\documentclass[aps,letterpaper,preprint,showpacs]{revtex4}

\usepackage{graphicx}

\renewcommand{\deg}{$^\circ$}

\begin{document}

\title{Spin injection from the Heusler alloy Co$_2$MnGe into Al$_{0.1}$Ga$_{0.9}$As/GaAs heterostructures}
 
\author{X.Y. Dong}
\author{C. Adelmann}
\author{J.Q. Xie}
\author{C.J. Palmstr\o m}\email{palms001@umn.edu}
\affiliation{Department of Chemical Engineering and Materials Science, University of Minnesota, Minneapolis, MN 55455, USA}
\author{X. Lou}
\author{J. Strand}
\author{P.A. Crowell}
\affiliation{School of Physics and Astronomy, University of Minnesota, Minneapolis, MN 55455, USA}
\author{J.-P. Barnes}
\author{A.K. Petford-Long}
\affiliation{Department of Materials, University of Oxford, Oxford OX1 3PH, UK}

\begin{abstract}

Electrical spin injection from the Heusler alloy Co$_2$MnGe into a $p$-$i$-$n$ Al$_{0.1}$Ga$_{0.9}$As/GaAs light emitting diode is demonstrated. A maximum steady-state spin polarization of approximately 13\,\% at 2\,K is measured in two types of heterostructures. The injected spin polarization at 2\,K is calculated to be 27\,\% based on a calibration of the spin detector using Hanle effect measurements. Although the dependence on electrical bias conditions is qualitatively similar to Fe-based spin injection devices of the same design, the spin polarization injected from Co$_2$MnGe decays more rapidly with increasing temperature.

\end{abstract} 

\pacs{72.25.Hg, 72.25.Mk,75.50.Cc}

\maketitle

Recent progress in the injection of spin-polarized carriers into semiconductors has bolstered the emerging field of semiconductor spintronics. To date, efficient electrical spin injection into semiconductors has been demonstrated only from magnetic semiconductors \cite{Fiederling} and conventional ferromagnetic metals such as Fe \cite{Zhu,Hanbicki}. Recently, a unique class of materials, ferromagnetic Heusler alloys, has received renewed attention due to the fact that some of them, such as Co$_2$MnGe \cite{Ishida} and NiMnSb \cite{deGroot}, have been predicted to be half-metallic. Half-metals have a band structure with only one occupied set of spin states at the Fermi level ($E_F$), resulting in 100\,\% spin polarization at $E_F$. In addition, many of the Heusler alloys have high Curie temperatures ($T_C > 600$\,\deg C) along with large magnetic moments ($> 3.5 \mu_B$/formula unit) \cite{Webster}. Furthermore, their lattice constants are close to those of III--V semiconductors, which makes them ideal candidates for epitaxial contacts \cite{Ambrose,JWDong,XYDong,Lund}. There have, however, been relatively few attempts to measure the spin polarization in Heusler alloys \cite{Soulen,Schmalhorst,Cheng}. In cases such as the full Heusler alloy Co$_2$MnGe, the gap for minority spin states is predicted to be less than 200\,meV \cite{Picozzi1}. In principle, this makes the spin polarization extremely sensitive to interfacial effects \cite{Galanakis} as well as temperature \cite{Dowben}. These are important considerations in designing any realistic spin injection device. 

This letter reports on the demonstration of electrical spin injection from the Heusler alloy Co$_2$MnGe into Al$_{0.1}$Ga$_{0.9}$As/GaAs light emitting diode (LED) heterostructures grown by molecular-beam epitaxy (MBE) on $p^+$-GaAs (100) substrates. Two types of heterostructures utilizing a highly-doped Schottky contact as a tunnel injector were investigated \cite{Hanbicki}. Sample I uses a quantum well (QW) as an optical detector and consists of (in growth sequence) a 300\,nm $p$-GaAs buffer layer ($1\times 10^{17}/\text{cm}^3$) / 200\,nm $p$-Al$_{0.1}$Ga$_{0.9}$As ($1\times 10^{16}/\text{cm}^3$) / 25\,nm $i$-Al$_{0.1}$Ga$_{0.9}$As / 10\,nm $i$-GaAs (QW) / 25\,nm $i$-Al$_{0.1}$Ga$_{0.9}$As / 100\,nm $n$-Al$_{0.1}$Ga$_{0.9}$As ($1\times 10^{16}/\text{cm}^3$) / 15\,nm $n/n^+$-Al$_{0.1}$Ga$_{0.9}$As / 15\,nm $n^+$-Al$_{0.1}$Ga$_{0.9}$As ($5\times 10^{18}/\text{cm}^3$). Sample II differs from Sample I only in that the 10 nm $i$-GaAs QW is missing. The electroluminescence of this sample originates from recombination in the bulk $p^+$-GaAs substrate. 

\begin{figure}[tb]
\includegraphics[width=8cm,clip]{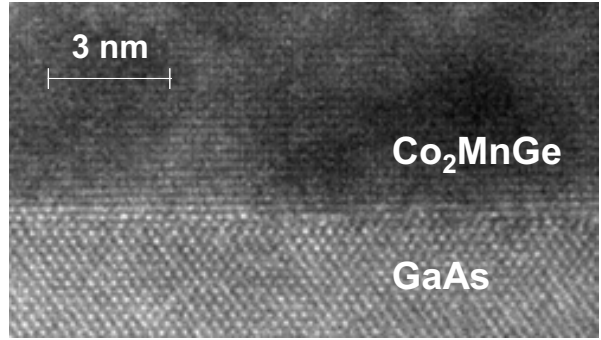}
\caption{High resolution cross-sectional TEM image of the Co$_2$MnGe/GaAs heterostructure.}
\end{figure}

After the growth of the LED structures, the samples were transferred under ultra-high vacuum ($< 10^{-10}$\,torr) to a second MBE system where a 7\,nm thick Co$_2$MnGe film was grown at 175\,\deg C, followed by a 2.5\,nm thick Al capping layer used to prevent oxidation in air. Streaky and sharp \emph{in-situ} reflection high-energy electron diffraction (RHEED) patterns were observed during Co$_2$MnGe growth, indicating that the films are single crystal with smooth surfaces. The spacing and intensities of the RHEED streaks suggest that Co$_2$MnGe grows in the (001) orientation and an L$2_1$-like crystal structure \cite{XYDong}. The high-resolution cross-sectional transmission electron microscopy (TEM) image in Fig.~1 indicates an abrupt coherent Co$_2$MnGe/GaAs interface. TEM diffraction shows that the film is L$2_1$-like with small disordered B2-like regions. X-ray diffraction revealed that the films are pseudomorphic with an out-of-plane lattice parameter of 5.86\,\AA .

Au contacts were deposited by a shadow mask technique, and then the samples were processed into 300\,$\mu$m diameter circular mesas using standard photolithography and both dry and wet chemical etching. The devices were then annealed in N$_2$ at 250\,\deg C for one hour. Electroluminescence (EL) measurements were carried out with a bias voltage applied between the substrate and the Co$_2$MnGe contact as shown in the inset of Fig.~2. Light was collected along the growth direction, which was parallel to the magnetic field. For the QW detector of sample I, the EL was dominated by recombination of heavy-hole excitons in the QW. The EL was measured for both circular polarizations and integrated over a window 3\,meV wide around the heavy-hole exciton peak. The EL polarization $P_{EL} = (I^+ - I^-)/(I^+ + I^-)$ was then calculated from the integrated intensities for right $(I^+)$ and left $(I^-)$ circularly polarized light. The EL spectrum for sample II was due to band-edge recombination in the $p^+$-GaAs substrate, and in this case the intensities were integrated over a window 40\,meV wide around the EL maximum. The electron spin polarization in the detector is $P_S = \alpha P_{EL} $, where $\alpha = 1$ for sample I and $\alpha = 2$ for the bulk detector of sample II \cite{OO}.

\begin{figure}[tb]
\includegraphics[width=8.5cm,clip]{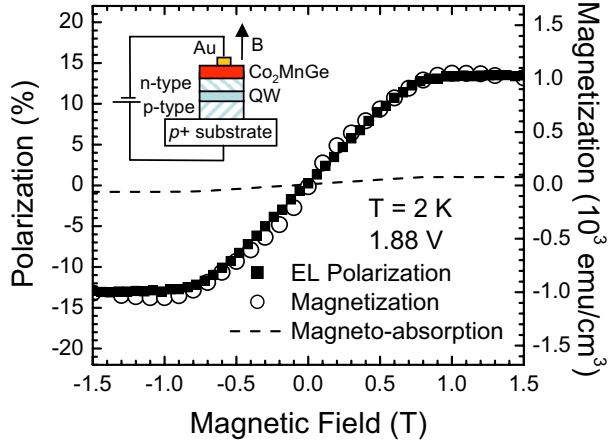}
\caption{Polarization signal $P_{EL}$ of sample I (squares) as well as magnetization (open circles) and magneto-absorption (dashed line) of the Co$_2$MnGe film as a function of applied magnetic field at 2\,K. The inset shows a schematic of the structure of sample I.}
\end{figure}

The EL polarization $P_{EL}$ measured for sample I at 2\,K and a bias of 1.88\,V is shown in Fig.~2 as a function of the applied magnetic field. The magnetization measured in the same geometry is also shown. As is evident from Fig.~2, the magnetization and the EL polarization show nearly identical magnetic field dependence. They both saturate at a field of 0.8\,T, above which $P_{EL}$ reaches a maximum value of approximately 14\,\%. The magneto-absorption of the Co$_2$MnGe film measured in a transmission experiment is shown as the dashed curve in Fig.~2 and is less than 1\,\%, approximately half the value for Fe films of comparable thickness. After subtraction of the magnetoabsorption from the raw data, the steady-state spin polarization in the QW is 13\,\% at 1\,T. 

\begin{figure}[tb]
\includegraphics[width=8cm,clip]{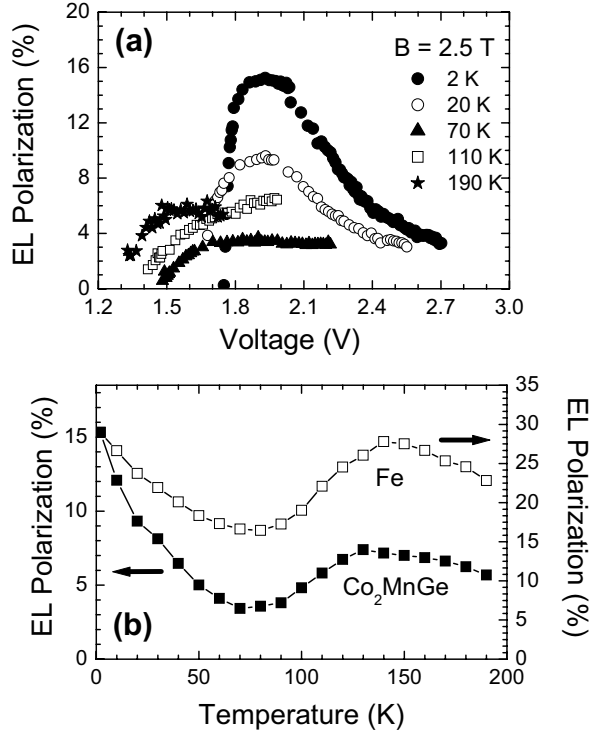}
\caption{\textbf{(a)} $P_{EL}$ for sample I as a function of the bias voltage in a field of 2.5\,T at the temperatures indicated in the legend. \textbf{(b)} The maximum value of $P_{EL}$ at each temperature is shown for Co$_2$MnGe (closed squares, left axis) and Fe (open squares, right axis).}
\end{figure}

The field dependence of $P_{EL}$ in Fig.~2, which follows the magnetization nearly exactly, is one of the explicit signatures of spin injection. A second distinguishing feature of spin injection into $p$-$i$-$n$ junction detectors is a marked dependence of $P_{EL}$ on the bias voltage and temperature, which influence the recombination and spin relaxation rates in the QW \cite{Adelmann}. EL polarization data measured for sample I as a function of bias at several different temperatures are shown in Fig.~3(a). The magnetic field for this set of measurements was fixed at 2.5\,T, which is above the saturation field of Co$_2$MnGe. The significant features of these data, including the maximum as a function of bias at low temperatures and the pronounced suppression of the signal near 70\,K, are similar to those found in bias-dependent measurements on Fe/(Al,Ga)As spin LEDs \cite{Adelmann}. The spin detector used in these measurements is of the same design as in Ref.~\onlinecite{Adelmann}, in which the bias dependence of the $P_{EL}$ was demonstrated to depend strongly on the recombination and spin relaxation times in the quantum well. Additional evidence that the detector plays a critical role can be seen in Fig.~3(b), which shows the \emph{maximum} polarization signal measured at each temperature for both Co$_2$MnGe and Fe. The minimum in $P_{EL}$ at 70\,K and the maximum near 150\,K appear for both injector materials. As discussed in Ref.~\onlinecite{Adelmann}, the existence of the minimum at 70\,K is due to a crossover from a low-temperature regime in which excitonic effects dominate to a high-temperature regime in which the  electrons in the QW are essentially free, and the spin relaxation rate is reduced. Although the magnitude of the signal in Co$_2$MnGe is smaller, the overall behavior observed in Fig.~3(b) is very similar for the two materials. The two important differences are the smaller overall signal for Co$_2$MnGe (15\,\% as opposed to 28\,\% for Fe) as well as the stronger decrease in $P_{EL}$ between 2 and 70\,K. 

Given the similarities between the data obtained using Fe and Co$_2$MnGe injectors, it is reasonable to ask whether the differences observed at the lowest temperatures in Fig.~3(b) are significant. Addressing this question requires a calibration of the semiconductor spin detector, which is accomplished by measuring the spin detection efficiency $\eta = 1/(1 + \tau_r/\tau_s)$, where $\tau_r$ and $\tau_s$ are the recombination and spin relaxation times for electrons in the QW \cite{OO}. Once $\eta$ is known, the \emph{injected} spin polarization $P_i = P_S/\eta$ can be calculated. In principle, $\eta$ can be determined from the optical Hanle effect in a transverse magnetic field \cite{Motsnyi}. In practice, the recombination time for QW detectors increases rapidly with temperature and depends strongly on bias, making a reliable calibration difficult. However, for bulk detectors such as sample II, the full Hanle curve can be measured up to room temperature, therefore allowing for a reliable determination of $\eta$ at all temperatures. The measured spin polarization $P_S$ for sample II (corrected for magnetoabsorption) is shown in Fig.~4 using solid squares, and the injected polarization $P_i$ deduced from the Hanle calibration is shown using solid circles. For comparison, $P_i$ obtained with an Fe injector on an otherwise identical device is shown using open triangles. The injected spin polarization for Co$_2$MnGe reaches a maximum of 27\,\% at 2\,K, in contrast to the value of 40\,\% reached with an Fe injector. More significantly, as suggested by the raw polarization measurements in Fig.~3(b), the spin polarization injected from Co$_2$MnGe decreases more rapidly with increasing temperature than in Fe. In the case of Fe, the injected polarization is approximately 15\,\% at room temperature. In contrast, the injected polarization at 300\,K is negligible for the Co$_2$MnGe device, in spite of the fact that room temperature is still well below the Curie temperature of 905\,K \cite{Cheng}.

\begin{figure}[tb]
\includegraphics[width=8cm,clip]{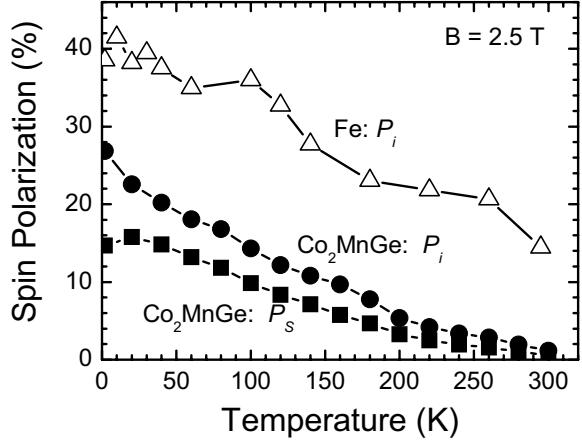}
\caption{The spin polarization $P_S$ (closed squares) for sample II is shown as a function of temperature along with the injected polarization $P_i$ (closed circles) determined using the calibration procedure described in the text. The injected polarization for Fe measured using an identical bulk spin detector is shown with open triangles.}
\end{figure}

The injected polarization for Co$_2$MnGe is therefore significantly below the value of 100\,\% that would be expected for a half-metal and shows stronger temperature dependence than observed for spin injection from Fe. Although the interpretation of absolute polarization measurements made using a spin-LED is subject to challenge, the injected polarization measured for Co$_2$MnGe is smaller than that for Fe as determined using \emph{both} QW and bulk detectors. Given the small gap ($\sim 200$\,meV) predicted for minority spins in Co$_2$MnGe \cite{Picozzi1} and some evidence for disordered (B2-like) regions in TEM, the apparent absence of half-metallic behavior is not too surprising and is consistent with earlier conclusions based on point-contact Andreev spectroscopy of bulk samples \cite{Cheng}.  Heusler alloys with a larger minority spin gap, such as Co$_2$MnSi \cite{Ishida,Schmalhorst}, may be more effective injectors.  Furthermore, the spin injection experiment described in this letter probes the polarization at the interface between a thin film of Co$_2$MnGe and Al$_{0.1}$Ga$_{0.9}$As. In addition to considering the electronic structure of the interface \cite{Galanakis}, a realistic theory will also have to incorporate the presence of alloy disorder in the film as well as the effects of non-zero temperature.  These factors will play an essential role in interpreting spin injection measurements on new materials.

This work was supported in part by the DARPA SPINS program, ONR, and the University of Minnesota MRSEC (NSF DMR-0212032).

\end{document}